# Large Language Models and Stock Investing: Is the Human Factor Required?

Ricardo Crisóstomo[a] and Diana Mykhalyuk[b]

## Abstract

This paper investigates whether large language models (LLMs) can generate reliable stock market predictions. We evaluate four state-of-the-art models—ChatGPT, Gemini, DeepSeek, and Perplexity—across three prompting strategies: a naïve query, a structured approach, and chain-of-thought reasoning. Our results show that LLM-generated recommendations are hindered by recurring reasoning failures, including financial misconceptions, carryover errors, and reliance on outdated or hallucinated information. When appropriately guided and supervised, LLMs demonstrate the capacity to outperform the market, but realizing LLMs' full potential requires substantial human oversight. We also find that grounding stock recommendations in official regulatory filings increases their forecasting accuracy. Overall, our findings underscore the need for robust safeguards and validation when deploying LLMs in financial markets.

[a] Corresponding author. Comisión Nacional del Mercado de Valores (CNMV), Edison 4, 28006 Madrid.
Email: rcayala@cnmv.es
[b] Comisión Nacional del Mercado de Valores (CNMV), Edison 4, 28006 Madrid.
Email: diana.mykhalyuk@cnmv.es

# 1. Introduction

The rapid evolution and increasing accessibility of large language models (LLMs) are opening new frontiers in financial markets. Modern LLMs are increasingly used to interpret complex financial documents, identify market trends, extract sentiment, or generate investment recommendations. However, these advanced capabilities raise a critical question: *Can LLMs autonomously generate reliable stock market predictions?*

Traditionally, in-depth market analysis has been the domain of investment professionals equipped with expert knowledge, real-time market data and specialized analytical tools. But recent advances in AI systems—including web browsing and real-time data retrieval—are democratizing financial intelligence[1]. By accessing and synthesizing live market data, breaking news and sentiment signals, LLMs are increasingly well positioned to support forward-looking investment strategies[2]. This technological shift has the potential to transform the investment landscape, empowering individuals with analytical resources that mirror those of financial professionals.

However, the increased adoption of LLMs in financial markets raises important concerns. As language models, LLMs are designed to generate fluent and confident answers, even when grounded in flawed reasoning, incomplete information or computational inaccuracies. Consequently, investors lacking financial expertise or critical judgment may place undue trust in AI recommendations, using model signals without properly understanding their risks and limitations.

Crucially, the extent to which LLMs can effectively access, synthesize, and reason over real-time financial information remains insufficiently understood. Web-based information is characterized by heterogeneous reporting standards, conflicting narratives and data sources, and potentially misleading content. These informational nuances introduce noise and ambiguity into LLMs reasoning, challenging models' ability to extract coherent signals and generate reliable investment recommendations.

[1] Web-browsing capabilities were progressively rolled out across major LLM platforms between late 2024 and early 2025.

[2] In financial markets, where asset prices can substantially change in response to emerging news and evolving expectations, access to timely, accurate information is essential for informed decision-making.

## Research Contributions

This paper provides a comprehensive evaluation of the current capabilities of modern LLMs in financial markets. By examining how different models integrate live market data and perform complex financial reasoning, our study makes five contributions to the literature:

i) **Prompt engineering and human oversight**. We assess the extent to which prompt design and human guidance influence the reliability of equity recommendations. Our analysis compares three prompting strategies. First, a naïve conversational approach, where models respond to simple user queries. Second, a structured framework that incorporates contextual inputs and calculation steps to support model reasoning. Third, chain-of-thought reasoning, in which human oversight is iteratively applied to ensure logical and coherent outputs.

ii) **Use of real-time information.** We evaluate how LLMs process and navigate real-time information flows. In particular, we assess whether models can effectively leverage dynamic inputs—such as breaking news, market prices, and regulatory disclosures—and how they manage challenges arising from potentially inconsistent, outdated, or conflicting data sources.

iii) **Benchmarking of LLM platforms**. We perform a systematic comparison of four leading LLMs—ChatGPT, Gemini, DeepSeek, and Perplexity—, assessing both their reasoning quality and quantitative investment performance. The benchmarking analysis covers a ten-month live evaluation window, capturing model performance across varying market conditions and software iterations.

iv) **Regulatory filings.** We examine whether grounding LLMs recommendations in official supervisory filing enhances predictive accuracy. By assessing the use of standardized regulatory disclosures, we evaluate the role that supervised, high integrity data can play in the emerging era of AI-driven information and decision-making.

v) **Taxonomy of reasoning failures.** Finally, we analyze the behavior of different models and prompting strategies in complex tasks requiring domain-specific expertise, contextual judgment and multi-step computations. This allows us to develop a taxonomy of reasoning failures and propose practical mechanisms for detecting, mitigating and managing such failures in real-world workflows.

## 2. Related Literature

LLMs are increasingly used in finance, with applications ranging from sentiment analysis to earnings prediction and ESG scoring (Zhang et al., 2018; Araci, 2019; Sokolov et al., 2021). Their strength lies in processing high-dimensional, unstructured textual data—offering an informational advantage over traditional econometric models that overlook linguistic nuance and contextual signals (Hwang & Kim, 2017).

Early applications of natural language processing (NLP) relied on dictionary-based approaches and rule-based sentiment systems (e.g., Antweiler & Frank, 2004; Loughran & McDonald, 2011). Subsequent advances introduced deep learning architectures, including BERT-style models applied to financial text mining and prediction tasks (Araci, 2019; Liu et al., 2020). Araci (2019) shows that transformer-based architectures extract signals from financial narratives with higher precision than previous methods. Similarly, Ke et al. (2019) document the economic value of textual information, demonstrating that the tone and context of corporate disclosures are predictive of future returns.

Building on these foundations, recent empirical assessments of generative LLMs in high-stakes financial tasks have yielded divergent results. While Pelster & Val (2024) find that ChatGPT's stock recommendations lack consistency and are sensitive to prompt design, Lopez-Lira & Tang (2023) demonstrate that ChatGPT-based scores can outperform traditional providers in return predictability tests. Kim et al. (2024) show that LLMs can extract valuable insights from financial statements—particularly when combined with structured inputs—while Gupta et al. (2023) demonstrate that grounding LLMs in real-time disclosures improves forecasting accuracy. Moreover, Cao et al. (2023) emphasize that Human–AI collaboration in equity analysis often outperforms either approach in isolation, highlighting the role of human judgment in complex decision-making.

Despite these advances, important methodological and interpretative challenges remain. Designed to prioritize linguistic plausibility over factual correctness, LLMs can produce fluent yet misleading outputs (Bubeck et al., 2023; Ji et al., 2023). Wu et al. (2023) document hallucinations in financial summaries, while Gilson et al. (2022) and Peng et al. (2023) report failures in math- and logic-intensive tasks. These limitations can be particularly relevant in asset-pricing, where small analytical errors can propagate through complex valuation frameworks and materially affect financial decisions.

## 3. Methodology

We employ a robust, out-of-sample investment framework to evaluate whether LLMs can generate reliable investment predictions. Specifically, we instruct four leading LLM platforms—ChatGPT, Gemini, DeepSeek and Perplexity—to generate predictive signals for equities expected to outperform the market, using prompts with varying degrees of human intervention. We further examine whether grounding the analysis in official regulatory filings enhances the quality and performance of LLM recommendations. All queries follow a standardized protocol that is consistently applied across platforms and evaluation dates.

### 3.1 Experimental Setup

Our evaluation is conducted over a live ten-month period, spanning April 2025 to January 2026. To replicate real-world investment conditions, we impose strict information boundaries: queries are submitted on the first trading day of each month, with the information cutoff set at the end of the preceding month. For each stock under consideration, we instruct the LLMs to compute a model-based outperformance score normalized to the [0, 1] interval, using three prompting strategies.

To ensure independence across periods, monthly prompts are initiated in fresh conversational contexts, limiting memory carryover and cumulative biases. Consistent with the information boundaries, outperformance scores are generated using the software iteration for each LLM available on the first day of the month. This ensures that model architectures and information sets are contemporaneously aligned, mirroring the real-time constraints faced by market participants. Annex 1 reports the LLM versions used throughout the evaluation period.

### 3.2 Prompt Design

We implement three prompting strategies with varying levels of analytical structure and human oversight: (i) a naïve query, (ii) a structured prompt, and (iii) chain-of-thought reasoning. These strategies are designed to capture a broad range of investors, spanning from casual retail investors to experienced professionals capable of actively guiding and refining model outputs. The prompt templates used in the empirical evaluation are provided in Annex 2.

## Naïve Query

The naïve query employs a simple, unstructured prompt that approximates how casual investors typically interact with conversational LLMs. The model is instructed to act as a financial manager and assign each stock a [0, 1] outperformance score based on its expected one-month return[3]. No guidance is provided regarding the underlying analytical framework, financial metrics, or weighting scheme.

Consequently, LLMs are free to determine which variables, evaluation criteria, and aggregation rules to apply, based entirely on their internal reasoning. This prompt specification captures the baseline performance of LLMs—i.e., their capacity to generate reliable equity predictions in absence of structured guidance, external grounding, or human oversight.

## Structured Approach

The structured prompt introduces analytical guidance and contextual constraints in line with financial analysis practices[4]. LLMs are instructed to compute outperformance scores based on six drivers of stock returns: (i) valuation, (ii) growth potential, (iii) financial health, (iv) technical indicators, (v) macroeconomic and sector risk, and (iv) sentiment. For each driver category, models are required to compute two commonly used financial metrics, applying predetermined weights to aggregate component scores into an overall [0, 1] outperformance score.[5]

Models are instructed to find the most recent data within the information cutoff and to incorporate forward-looking expectations, if available. This prompt design evaluates the model's ability to follow precise instructions, execute multi-step analytical reasoning, process heterogeneous financial information, and synthesize intermediate outputs into coherent financial recommendations. Table 1 summarizes the multi-factor scoring framework, including driver categories, financial metrics, weights, and their economic rationale.

---

[3] The scoring system facilitates a systematic comparison of LLM-generated recommendations, enabling a comprehensive evaluation of model outputs across platforms and prompt strategies.

[4] Prompting instructions adhere to the guiding principles for output coherence and reasoning accuracy established in Bsharat et al. (2023).

[5] To facilitate aggregation and cross-section comparability, models are instructed to normalize financial metrics into a [0,1] interval based on the empirical distribution of values within the IBEX 35 index.

## Chain-of-Thought Reasoning

CoT reasoning incorporates explicit human oversight through iterative, dialogue-based interactions. Starting from the structured-prompt output, model responses are individually reviewed and refined across successive prompts, correcting inconsistencies in models' reasoning, such as computational errors, reliance on stale or outdated information, and misinterpretation of financial metrics.

CoT reasoning provides an upper-bound estimate of the attainable LLM performance, illustrating the extent to which model's outputs can be improved through active human intervention[6].

**Table 1. Multi-Factor Scoring Framework**

| Category | Weight | Metric (sub-weight) | Rationale |
|---|---|---|---|
| **Valuation** | 20% | **P/E Ratio** (60%) | Price to Earnings ratio. Lower values indicate undervaluation (high earnings relative to market price). |
| | | **P/B Ratio** (40%) | Price to Book ratio. Lower values indicate undervaluation (high book value relative to market price). |
| **Growth Potential** | 20% | **EPS Growth** (60%) | Earnings Per Share growth. Higher values indicate robust bottom-line earnings expansion. |
| | | **Revenue Growth** (40%) | Higher values indicate robust top-line revenue expansion. |
| **Financial Health** | 15% | **Debt/Equity** (60%) | Lower values indicate low financial leverage and indebtedness. |
| | | **ROE** (40%) | Return on equity. Higher ROE indicates efficient capital use (high return relative to equity). |
| **Technical** | 15% | **Momentum** (60%) | Measures recent price performance. |
| | | **RSI** (40%) | Relative Strength Index. Identifies overbought/oversold conditions. |
| **Macro & Sector** | 15% | **Industry Growth** (60%) | Captures sectoral growth. |
| | | **Sector Outlook** (40%) | Assesses risk and opportunity at the sector level. |
| **Sentiment** | 15% | **News Sentiment** (80%) | Measures recent news sentiment. |
| | | **Analyst views** (20%) | Gauges professional market recommendations. |

[6] Naïve and structured queries are executed on the first trading day of each month. CoT queries may extend beyond that date due to their iterative nature, relying exclusively on information available up to the cutoff date.

## 3.3 Evaluation Framework

We implement a dual evaluation framework that assesses both the reasoning integrity and the financial performance of LLM investment predictions.

### 3.3.1 Reasoning Quality

We first evaluate the reasoning integrity of LLM outputs. This assessment focuses on distinguishing between superficial textual fluency and analytical competence. This distinction is critical, as responses that appear plausible but rely on flawed logic, incorrect computations or fabricated inputs possess no intrinsic informational value.

Our evaluation examines whether LLMs exhibit systematic reasoning failures, including logical inconsistencies, arithmetic errors, and reliance on stale or outdated information. We also assess models' ability to correctly compute and interpret financial and valuation risk metrics, a prerequisite for sound financial reasoning.

By focusing on how each model constructs, computes, and justifies its predictions, this evaluation isolates the reasoning process from subsequent market realizations. Accordingly, our analysis decouples reasoning quality from realized returns, allowing us to assess whether any observed performance reflects genuine analytical competence rather than *ex-post* coincidence with market movements.

### 3.3.2 Economic Performance

After assessing LLM's reasoning quality, we proceed to evaluate the quantitative performance of the model-generated investment signals.

#### Trading strategy

For each LLM platform and prompt design, we construct a monthly long-short portfolio based on model-generated outperformance scores. Positions are initiated on the first trading day of each month, and held until the final trading day. This procedure yields ten independent monthly investment cycles spanning April 2025 through January 2026. At the beginning of each month, portfolios are fully rebalanced to reflect the new information and trading signals obtained from each platform and prompt strategy. Table 2 summarizes the trading strategy and rebalancing procedure.

**Table 2. Trading Strategy and Rebalancing Procedure**

| | |
|---|---|
| **Investment Universe** | All 35 stocks included in the IBEX-35 index. |
| **Portfolio Construction** | Long the five stocks with the highest outperformance scores; short the five stocks with the lowest outperformance scores. Equally weighted portfolio. |
| **Holding Period** | Monthly. Investment positions are initiated on the first trading day and held until the last trading day of the month. |
| **Rebalancing** | Portfolios are rebalanced on the first trading day of each month. |
| **Evaluation Horizon** | April 2025 – January 2026; ten independent monthly cycles. |
| **Benchmark** | IBEX-35 Index (total return). |

## Ex-ante Evaluation Metrics

To evaluate the performance of LLM-generated signals, we employ four complementary metrics that jointly capture *excess return, risk-adjusted performance,* and *predictive accuracy:*

- **Excess return ($\alpha$):** Measures the excess return of the LLM-based portfolio relative to the benchmark. This metric isolates the incremental contribution of LLM stock selection from general market movements. Returns for both the benchmark and individual constituents are computed on a total return basis, incorporating dividends and price appreciation.

- **Information ratio ($IR$):** Quantifies the risk-adjusted return of the portfolio. It provides a standardized measure of whether excess performance is achieved without disproportionate exposure to volatility. The $IR$ is defined as the ratio of the mean monthly excess return to the standard deviation of that excess return (tracking error).

- **Directional accuracy:** Computes the proportion of correct directional predictions across the full cross-section of assets. An outperformance (underperformance) prediction is classified as correct if the realized total return of the stock exceeds (falls below) the return of the benchmark over the corresponding period.

- **Weighted F1-score.** Represents the harmonic mean of precision and recall, providing a balanced assessment of predictive reliability. By penalizing both false positives and false negatives, the F1-score evaluates model's consistency in identifying both outperformers and underperformers. To capture classification performance across both positive and negative classes, we compute class-specific F1-scores for the outperformance and underperformance categories and report their average.

A central feature of our methodology is that all LLM outperformance scores are generated on a strictly *ex-ante* basis. For each month *t*, scores are obtained at the beginning of each period using an information set restricted to the close of the preceding month *t-1*. Since LLM predictions are recorded in real-time, this process occurs without any knowledge of subsequent market realizations—which had not yet materialized at the time of the query. This protocol ensures a genuine out-of-sample evaluation, eliminates look-ahead bias, and preserves the temporal integrity of the research design.

## 3.4 Regulatory Filings

In addition to the naïve, structured and CoT queries, we evaluate whether incorporating official regulatory filings can improve the quality and performance of LLM investment recommendations. Regulatory-mandated disclosures—designed to foster transparency, standardization and comparability—are particularly well suited to examine the role that structured, high-integrity data can play in the emerging era of AI-driven information processing and decision-making.

For each company and monthly query, we upload the official regulatory filings released during the preceding month. Regulatory disclosures are obtained directly from "*Privileged Information"* and "*Other Relevant Information"* repositories of the Spanish securities market supervisor (CNMV)[7]. Along with *ad hoc* disclosures and corporate communications, regulatory filings include annual and semi-annual financial reports, providing LLMs with timely, standardized, and comparable information for the stocks under consideration[8].

[7] The CNMV is the competent authority for securities listed on Spanish markets, including constituents of the IBEX 35. Corporate disclosures are retrieved from official filings published on the CNMV webpage (www.cnmv.es)

[8] As of April 1, 2025, the free-tier version of Gemini did not support attachment processing. Consequently, regulatory filings are analyzed using ChatGPT, Perplexity, and DeepSeek.

# 4. Results and Discussion

We now present the findings from our systematic evaluation of LLM platforms. Our analysis examines both the integrity of models reasoning and the performance of trading portfolios constructed from model-generated signals.

## 4.1 Assessment of Reasoning Quality

The integrity of LLM reasoning is a prerequisite to generate reliable investment recommendations. In financial markets—where asset prices respond rapidly to breaking news, data revisions, and shifts in sentiment—internal coherence, temporal validity, and numerical consistency are essential. When financial recommendations are grounded on stale information, misinterpreted ratios, or erroneous computations, the resulting analyses become unreliable and can be materially misleading.

Nevertheless, it is important to recognize the inherent complexity of the required task. Under naïve prompts, LLMs are effectively asked to generate a predictive framework for financial markets—an undertaking widely regarded as particularly challenging in economics. Under the structured and CoT prompts, models must retrieve live, time-sensitive inputs, compute and normalize financial metrics (often across accounting regimes and currencies), apply predefined weighting schemes, and accurately interpret domain-specific instructions. Taken together, these requirements constitute a stringent test of LLM reasoning capabilities, closely approximating the analytical workflow performed by equity analysts in real-world settings.

Overall, our assessment shows that LLMs produce coherent and persuasive narratives; however, beneath this linguistic fluency, we observe recurring reasoning and computational failures. The limitations of current LLMs manifest in inconsistencies in retrieved data, misinterpretation of financial figures, computational mistakes, and meta-reasoning failures—yielding a significant gap between textual plausibility and factual accuracy. Although failures are not omnipresent, and many tasks are executed correctly, the recurrence of reasoning errors across models, prompt strategies and testing periods, raises significant concerns about the autonomous use of LLMs in financial decision-making. Table 3 presents a taxonomy of LLM's reasoning errors, which we classify into four categories:[9]

[9] We do not report quantitative error rates as such metrics are highly dependent on specific prompt designs, model architectures and the probabilistic nature of language models.

### A. Live Data Retrieval

A persistent limitation concerns the factual integrity of retrieved data. When prompted to access live, dynamic or time-sensitive sources, LLM's outputs recurrently contain outdated, inconsistent, or fabricated figures. These inaccuracies often stem from confusion over table structures, conflation of outdated and newly retrieved content or reuse of stale information from previous queries, leading to outputs that mix fiscal periods, currencies or even data from different firms, as if they were directly comparable.

### B. Financial Interpretation

A second class of errors involves semantic failures in financial reasoning. A recurring issue is the incorrect interpretation of fundamental ratios. For instance, LLMs incorrectly treat lower values of Price-to-Book (P/B), Price-to-Earnings (P/E), or Debt-to-Equity ratios as negative, failing to recognize that such values indicate undervaluation or lower leverage. Similarly, in the analysis of multinational firms, models may conflate figures reported in different currencies or accounting regimes, producing metrics that mislead peer comparisons.

### C. Computational Processing

Even when data retrieval and financial interpretation are sound, quantitative execution may fail. A prominent issue arises in the weighting and aggregation of financial metrics. In multiple instances, category scores—defined as the average of two underlying indicators—fell outside the mathematically feasible range implied by those indicators. Additional failures include inappropriate imputations of missing values, arithmetic inconsistencies in weighted sums, and omission or incorrect propagation of intermediate results in multi-stage computations.

### D. Meta-Reasoning and Transparency

The fourth category encompasses higher-order reasoning failures, including limitations in introspection, auditability, and complexity management. When prompted to revise errors, models often provide confident yet flawed corrections that preserve the original mistake, suggesting limited self-evaluation capabilities. In other instances, models exhibit opaque reasoning, failing to provide clear justification, rationale, or weightings for outperformance scores. Similarly, when processing documents, models display inconsistent reading behavior, failing to digest the entire document—sometimes without disclosure—or answering certain queries immediately while spending several minutes on comparable prompts.

**Table 3: Taxonomy of LLM Reasoning Failures**

| Error Type | Description | Empirical Example |
|---|---|---|
| **A. Web Browsing and Data Retrieval**<br>Errors arising from LLM interaction with external or real-time data sources. | | |
| **A1. Entity Mismatch** | Retrieves data for the wrong company. | Uses financial statements for Amadeus Fire AG instead of Amadeus IT Group S.A. |
| **A2. Fabricated Data** | Generates plausible but non-existent figures. | Uses a closing price of 18.40 when no such value appears in available data. |
| **A3. Temporal Mismatch** | Combines figures from non-comparable periods. | Uses 2025 EPS together with stock price from 2024. |
| **A4. Table Misinterpretation** | Misread rows or columns in structured web tables. | Extracts incorrect P/E of 18.34, confusing rows in a structured web table. |
| **A5. Search Instability** | Displays inconsistent search behavior. | Consults dozens of sources in one query; only two in a comparable request. |
| **B. Financial Misinterpretation**<br>Conceptual errors in financial reasoning. | | |
| **B1. Financial Misunderstanding** | Erroneous interpretation of financial ratios. | Interprets low positive P/B or P/E as negative rather than potential undervaluation. |
| **B2. Overconfident Summation** | Converts ambiguous evidence into precise numerical claims. | Infers a P/E ratio of 22x from a narrative describing the firm as "overvalued" |
| **B3. Formula Misapplication** | Incorrectly applies financial formulas | Uses a single absolute figure to evaluate EPS growth. |
| **B4. Currency Inconsistency** | Combines figures denominated in different currencies. | Aggregates USD-denominated debt with EUR-based market capitalization. |
| **B5. Distorted Normalization** | Applies normalization schemes that undermine comparability. | Normalizes firm scores differently across comparable queries. |
| **B6. Inappropriate Input** | Uses indicators that do not properly represent the intended risk category. | Uses country-level data to proxy the overall risk of a multinational firm. |

**Table 3: Taxonomy of LLM Reasoning Failures (continued)**

| Error Type | Description | Empirical Example |
|---|---|---|
| **C. Computational Processing**<br>Errors in arithmetic execution, aggregation logic, or multi-step numerical reasoning. | | |
| **C1. Aggregation Error** | Miscalculates weighted sums or averages. | Produces a weighted score of 0.71 when correct output is 0.56. |
| **C2. Carryover Error** | Reuses stale output values from prior responses. | Repeats a previous final score despite showing different calculations. |
| **C3. Multi-Step Calculation Failure** | Omits or simplifies intermediate computations. | Fails to calculate RSI despite having the required data. |
| **C4. Missing-Value Distortion** | Substitutes missing or extreme values with arbitrary figures. | Replaces missing earnings with zero and uses it in computations. |
| **C5. Output Instability** | Produces inconsistent outputs for identical prompts. | Computes score of 0.75 that is subsequently changed in an equivalent prompt. |
| **D. Meta-Reasoning and Transparency**<br>Higher-order reasoning failures in consistency, clarity and auditability. | | |
| **D1. Failed Self-Correction** | Acknowledges an error but continues using incorrect value. | Corrects a company's ROE verbally but still uses erroneous figure. |
| **D2. Complexity Reduction Bias** | Ignores constraints or dependencies in multi-asset or multi-step tasks. | Performs full calculations for some firms but simplifies calculations in others, affecting comparability. |
| **D3. Default Output Substitution** | Uses generic outputs instead of firm-specific calculations. | Assigns a default score of 0.85 to all banks without differentiation. |
| **D4. Attachment Processing Inconsistency** | Applies inconsistent depth when processing attachments. | Reads only a small portion of an attachment in one query; processes the full document in an equivalent request. |
| **D5. Arbitrary Uncertainty Adjustment** | Applies non-standard numerical adjustments without methodological justification. | Discounts a computed score from 0.58 to 0.55 to reflect "uncertainty." |
| **D6. Opaque Reasoning** | Fails to disclose the weighting or scoring logic underlying outputs. | Produces a score of 0.72 but cannot reproduce the underlying factors or weights. |

## 4.2 Managing Reasoning Failures

Our findings demonstrate a critical divergence: although large language models exhibit exceptional linguistic fluency, they lack robustness in financial reasoning. Because textual explanations appear coherent even when numerical outputs are incorrect, LLMs mistakes often go undetected, creating a false impression of precision. This phenomenon highlights a *fluency trap*, where rhetorical credibility does not imply analytical integrity.

### Model Progress (April 2025 – January 2026)

Across our evaluation window, model capabilities evolved substantially. In April 2025, LLMs were in the early stages of integrating web-browsing and live data retrieval tools, and exhibited limited financial reasoning. Recent model iterations demonstrate improved arithmetic accuracy, lower hallucination rates, better handling of missing values, and greater transparency in underlying assumptions. Moreover, higher-order reasoning capabilities, including financial understanding, internal consistency, and auditability, have also improved significantly, leading to a reduction in reasoning failures.

Yet, despite these recent advances, as of January 2026 recurring limitations persist in data retrieval, computational processing, financial interpretation, and meta-reasoning. Taken together, these findings suggest that LLMs have not yet achieved the robustness required for unmonitored deployment in high-stakes financial environments.

### Mitigation Strategies

To address the reasoning failures summarized in Table 3—spanning data retrieval, financial interpretation, computational processing, and meta-reasoning—we propose four practical mitigation principles for LLM-driven quantitative workflows. Although institutional applications may require more formal governance frameworks, these principles can substantially reduce the incidence of reasoning errors and provide safeguards against the stochastic nature of LLM outputs.

1. **Enforce *show your work*:** Prompts should require models to explicitly articulate reasoning paths, assumptions, and intermediate calculations before presenting final outputs. Enforcing a *show your work* discipline encourages LLMs to perform structured reasoning rather than converge prematurely on unsupported conclusions. This reduces the likelihood of calculation hallucinations—where plausible numbers are generated without proper computation—and ensures that all required inputs are retrieved and displayed prior to aggregation. Moreover, this transparency facilitates human oversight by exposing inconsistent formulas, logical errors, or omitted variables.

2. **Verify data provenance:** Models should be required to provide explicit data citations (e.g., URLs, document titles, publication dates, or table references) for all numerical inputs. Anchoring quantitative claims to verifiable sources mitigates the risk of relying on outdated, misinterpreted, or fabricated information, enabling traceability of data lineage. However, citations themselves must be also independently verified, as LLMs may hallucinate plausible but nonexistent references or misread legitimate documents.

3. **Perform iterative validation:** Rather than accepting initial outputs, users should incorporate explicit validation routines. Through recursive prompting or automated checks, LLMs can be tasked with reviewing the internal consistency and accuracy of their outputs, verifying numerical ranges, bounded quantities, and mathematical constraints. This second-pass evaluation shifts the model from a purely generative role toward an evaluative one, frequently uncovering errors such as impossible averages, inconsistent computations, or omitted intermediate steps.[10]

4. **Embed Human-in-the-Loop oversight:** Given the fluency trap—where confident language may obscure analytical deficiencies—human oversight remains a fundamental governance principle. Human-in-the-loop (HITL) supervision should operate as an overarching control throughout the entire analytical workflow. Beyond mechanical validation, expert oversight is essential to assess semantic and contextual coherence, verify metric definitions, reconcile discrepancies across conflicting data sources, and ensure the appropriate treatment of domain-specific terminology and non-recurring events.

[10] In practice, this validation can be partially automated. For instance, models can be instructed to verify arithmetic constraints, checks internal consistency and recompute quantitative outputs using independent execution engines, flagging divergences for further review.

## 4.3 Economic Performance of LLM-Generated Portfolios

### 4.3.1 Can LLMs Outperform the Market?

For each LLM–prompt configuration, we construct a ranked long–short portfolio that is rebalanced monthly from April 2025 through January 2026. Each month, the strategy takes long positions in the five IBEX-35 constituents with the highest *ex-ante* outperformance scores and short positions in the five with the lowest scores. Financial performance is evaluated in terms of excess return relative to the IBEX-35 benchmark and risk-adjusted performance, measured by the Information Ratio. Table 4 reports the performance metrics for each model–prompt combination.

**Naïve queries.** We begin by examining the predictive content of simple, conversational prompts. Across all LLM platforms, these queries generate an average monthly excess return of 0.35%, which is statistically indistinguishable from zero. Correspondingly, risk-adjusted performance is also weak, with information ratios well below levels typically regarded as meaningful in the asset-pricing literature (monthly IR = 0.06). These findings suggest that naïve queries—such as those typically posed by retail or uninformed investors (e.g., *"Which stocks should I buy?"*)—lack meaningful predictive power and fail to improve the returns or the risk profile of a passive benchmark strategy.

**Structured prompts.** We next examine whether guiding models with prompts that provide contextual inputs and analytical structure enhances the economic performance of LLM-generated portfolios. Our results show that anchoring models with domain-specific information—such as key financial indicators, instructions to use recent data, or weighting schemes—improves the coherence of model reasoning and reduces ambiguity and informational noise.

Under structured prompting, average excess return increases to 2.24% monthly, a substantial improvement compared to naïve queries. Notably, both Perplexity and Gemini achieve excess returns over the market benchmark that are statistically significant at 1% level. Anchoring models with explicit instructions also improves risk-adjusted performance, with aggregate IR rising to 0.58. Similarly, the average t-statistic increases to 2.16 compared to 0.18 in naïve prompting.

However, unsupervised structured queries exhibit notable cross-model heterogeneity. While Perplexity and Gemini achieve statistically significant alphas, ChatGPT and DeepSeek display weaker performance, with an average IR of 0.37 and insignificant excess returns ($t$-statistics of 1.41 and 1.22), respectively. This dispersion is consistent with the reasoning errors documented in Section 4.1. Since reasoning failures typically propagate into final outputs in unsupervised settings, autonomous LLM deployments generate an "instability of alpha" where certain model versions, time periods, or specific queries are adversely affected by financial inconsistencies or numerical miscalculations, leading to unstable performance.

**Chain-of-Thought reasoning.** The strongest performance arises when structured prompting is combined with iterative review and human oversight. Under CoT reasoning, trading portfolios achieve an average monthly excess return of 3.04% and a risk-adjusted Information Ratio of 0.68—the highest across all prompting strategies. Moreover, all four LLM platforms—Perplexity, Gemini, ChatGPT, and DeepSeek—generate statistically significant excess returns, with an average $t$-statistic of 2.45.

Because chain-of-thought prompting builds upon structured-query outputs, the incremental gains cannot be attributed solely to improvements in model reasoning. Rather, the enhanced performance under CoT reflects a collaborative Human–AI framework. By requiring models to externalize their reasoning processes—including explicit data references, intermediate calculations, and logical steps—analytical failures become transparent and can be systematically corrected.

Specifically, CoT enhancements operate through two complementary mechanisms. First, CoT prompts facilitate LLMs self-correction by requiring the explicit articulation of analytical steps; this reduces the likelihood that models bypass logical constraints or converge prematurely on unsupported conclusions. Second, increased transparency enables human supervisors to identify and correct financial inconsistencies, numerical mistakes, and data retrieval failures, preventing error propagation across subsequent queries.

### 4.3.2 Classification Accuracy and F1-Score

To supplement portfolio-level analytics, we evaluate whether LLMs correctly classify individual stocks across the full cross-section of assets. Rather than focusing on the extreme quintiles (the highest- and lowest-scoring stocks), this analysis assesses predictive accuracy across the entire IBEX 35 universe.

Table 4 reports the classification accuracy and F1-scores across models and prompt strategies. Naïve queries achieve an accuracy of 0.554 and an F1-score of 0.525, indicating modest but positive predictive content. Consistent with the portfolio-level results, these metrics suggest that zero-shot prompts extract signals that perform better than a random classifier[11]. However, the improvement over a random strategy is limited and remains vulnerable to reasoning failures, constraining the economic value of simple queries for investment decision.

[11] For binary classifications, a random classifier would yield an expected accuracy of 0.50.

Structured prompts improve average accuracy to 0.579 and raise the F1-score to 0.543. These gains confirm that explicit analytical guidance enhances models' ability to discriminate between outperforming and underperforming assets. However, substantial dispersion persists across models and time periods. For instance, over the ten-month live sample, directional accuracy ranges from 0.343 to 0.714 for ChatGPT and 0.371 to 0.714 for DeepSeek, underscoring performance heterogeneity in unsupervised queries.

The strongest classification performance emerges again under chain-of-thought prompting. While directional accuracy (0.571) is comparable to that of structured prompts, dispersion is lower, and the F1-score increases to 0.552. The improvement in F1-score reflects more effective handling of asymmetric classification errors—an important consideration in long-only strategies, where false positives (i.e., misclassifying underperforming stocks as outperformers) can be particularly costly. Under CoT prompting, the F1-score for identifying outperforming stocks increases by more than 6% relative to naïve queries and by 5% relative to structured unsupervised prompts.

**Prompt engineering summary.** Overall, our evidence from both trading portfolios and cross-sectional classification tests indicates that market outperformance is attainable using current LLM technologies. However, the persistence and propagation of reasoning errors materially compromise the reliability of autonomous deployments. In unsupervised settings, reasoning failures frequently remain undetected and translate directly into erroneous financial decisions, degrading performance and amplifying risk and instability.

Accordingly, our findings suggest that human oversight remains a critical component of robust financial applications. Rather than functioning as autonomous agents, LLMs appear most effective within supervised workflows, where expert judgment serves as a validation layer—mitigating model failures, and ensuring economically coherent outputs.

**Table 4: Economic Performance and Classification Accuracy**

| Model / Prompt | Naïve | Structured | CoT |
|---|---|---|---|
| **Panel A - Excess Return (monthly)** | | | |
| ChatGPT | -0.0072 | 0.015 | 0.030** |
| | (-0.585) | (1.406) | (2.020) |
| DeepSeek | 0.0069 | 0.024 | 0.027* |
| | (0.445) | (1.224) | (1.675) |
| Gemini | 0.0094 | 0.028*** | 0.030*** |
| | (0.618) | (2.894) | (3.120) |
| Perplexity | 0.0050 | 0.023*** | 0.035*** |
| | (0.263) | (3.129) | (3.007) |
| **Prompt strategy** | **0.0035** | **0.022** | **0.030** |
| **Panel B - Information Ratio (monthly)** | | | |
| ChatGPT | -0.15 | 0.41 | 0.61 |
| DeepSeek | 0.13 | 0.34 | 0.56 |
| Gemini | 0.18 | 0.86 | 0.89 |
| Perplexity | 0.083 | 0.70 | 0.65 |
| **Prompt strategy** | **0.060** | **0.58** | **0.68** |
| **Panel C - Accuracy** | | | |
| ChatGPT | 0.569 | 0.569 | 0.543 |
| DeepSeek | 0.549 | 0.569 | 0.549 |
| Gemini | 0.546 | 0.571 | 0.586 |
| Perplexity | 0.551 | 0.609 | 0.606 |
| **Prompt strategy** | **0.554** | **0.579** | **0.571** |
| **Panel D – F1 Score** | | | |
| ChatGPT | 0.527 | 0.524 | 0.528 |
| DeepSeek | 0.526 | 0.534 | 0.534 |
| Gemini | 0.525 | 0.546 | 0.567 |
| Perplexity | 0.521 | 0.568 | 0.579 |
| **Prompt strategy** | **0.525** | **0.543** | **0.552** |

*, **, and *** denote statistical significance at 10%, 5%, and 1% levels, respectively. *t*-statistics are reported in parentheses and are computed using Newey–West (1987) with one lag to account for heteroskedasticity and autocorrelation.

### 4.3.3 Benchmarking of LLM Platforms

We next compare the four LLM platforms across both portfolio returns and classification metrics. Table 5 summarizes the aggregate performance by LLM platform in terms of excess returns, Information Ratios, classification accuracy, and F1-scores.

Overall, **Perplexity** emerges as the strongest-performing model across most metrics. It achieves high average monthly excess returns under both structured and chain-of-thought prompting (2.3% and 3.5%, respectively), accompanied by strong Information Ratios (0.70 and 0.65). This financial outperformance is mirrored in cross-sectional classification results, where Perplexity records the highest accuracy (0.589) and F1-score (0.556). The consistency across portfolio returns and statistical classification suggests that Perplexity is particularly effective at extracting predictive signals and translating them into economically meaningful outcome. This advantage is consistent with its search-oriented architecture, which integrates retrieval-based mechanisms alongside language modeling, reducing reasoning failures and enhancing robustness across sophisticated prompt strategies.

**Gemini** ranks second overall, outperforming ChatGPT and DeepSeek in most dimensions. It delivers high excess returns, strong Information Ratios, and robust classification metrics. While it does not match Perplexity's peak performance in statistical classification, Gemini's performs particularly well under unsupervised configurations, where it achieves the highest returns and risk-adjusted performance. This reflects fewer failures in non-supervised queries and a reliable response to explicit analytical guidance.

**ChatGPT** ranks third overall and exhibits high sensitivity to prompt design. While structured and chain-of-though prompting yield solid economic performance, simple conversational queries generate negative excess returns and Information Ratios (see Table 4). In contrast, classification metrics remain competitive even under naïve prompting. This divergence suggests that, in absence of analytical guidance and human oversight, ChatGPT struggles to translate directional signals into consistent portfolio returns.

Finally, **DeepSeek** displays the most conservative performance profile. Although it generates positive excess returns across all prompt strategies, it consistently underperforms other platforms in risk-adjusted performance and classification quality. Notably, DeepSeek exhibits weaker results under CoT prompts—the setting in which other models achieve strongest gains. This pattern suggests that even with human-led reasoning improvements, DeepSeek does not easily achieve robust and reliable investment performance.

**Table 5: Performance Metrics by LLM Platform**

| Model | Excess Return | Information Ratio | Accuracy | F1-Score |
|---|---|---|---|---|
| ChatGPT | 0.013 | 0.29 | 0.560 | 0.526 |
| DeepSeek | 0.019 | 0.35 | 0.555 | 0.531 |
| Gemini | 0.022 | 0.64 | 0.568 | 0.546 |
| Perplexity | 0.021 | 0.48 | 0.589 | 0.556 |

#### 4.3.4 Impact of Regulatory Disclosures

We next examine whether incorporating regulatory filings in the investment process enhances the quality and performance of LLM-generated recommendations. Since regulatory disclosures are standardized, high-integrity, and comparatively low-noise information sources, they constitute a natural grounding mechanism for language models. Accordingly, such disclosures are expected to improve models reasoning by reducing ambiguity, enhancing cross-firm comparability, and constraining model reliance on incomplete or inconsistent narratives. [12]

Table 6 reports the performance metrics for portfolios constructed with and without regulatory filings. The regulatory-filings portfolio is implemented as a refinement of the firm-level signal, obtained by averaging the score derived from regulatory disclosures with the score computed under the naïve, structured and CoT prompts. Portfolios incorporating regulatory filings achieve improvements in risk-adjusted performance—most notably in Information Ratios—as well as higher classification accuracy and F1-scores. Notably, the IR of Perplexity-based queries increases by more than 17 percentage points, while that of ChatGPT-based queries rises by nearly 10 points. In addition, directional accuracy across models and prompting strategies improves from 0.568 to 0.579.

Because prompts lacking analytical structure and guidance are more vulnerable to reasoning failures, the inclusion of regulatory disclosures generates the largest gains in simple conversional queries. Empirically, grounding naïve prompts in regulatory filings increases monthly excess returns from 0.25% to 1.02% and the IR from 0.06 to 0.19.

[12] It is important to emphasize, however, that regulatory filings face inherent limitations for short-term investment horizons. Regulatory disclosures are often backward-looking—annual and semiannual reports are typically released with a one-to-two-month lag—and are not issued uniformly across all firms in every period. Consequently, it is challenging for filings alone to provide the timely or continuously updated signals that typically drive investment returns over short-term periods.

Similarly, directional accuracy improves from 0.554 to 0.579, while F1-scores increase from 0.525 to 0.553.

Regulatory disclosures also enhance performance under sophisticated prompting strategies. For the top model-prompt configuration—Perplexity with CoT prompting monthly excess returns increase from 3.5% to 4.0%, yielding a cumulative excess return of 40.2% over the benchmark in our ten-month window. Excluding DeepSeek's results—which reflect limitations in processing lengthy document attachments—all aggregate performance metrics improve when regulatory disclosures are integrated into the investment process[13]. These findings indicate that even for queries enhanced through structured reasoning and iterative supervision, grounding LLM predictions in standardized, high-integrity disclosures further strengthen the information content of model-generated signals.

In contrast, DeepSeek represents a notable exception. Contrary to the general pattern, its performance deteriorates when regulatory filings are incorporated. Diagnostic analysis suggests that this decline stems from limitations in processing lengthy attachments: DeepSeek frequently fails to parse complete filings, generating oversimplified summaries that do not reliably extract decision-relevant information. This divergence underscores an important governance implication: even when inputs are relevant and high quality, reliable LLM outputs are not guaranteed. Contextual grounding enhances informational value, but only if models can effectively process the underlying data, and if the empirical deployment is supported by rigorous validation and appropriate oversight.

**Table 6: Performance Metrics Including Regulatory Filings**

| Metric | Naïve | | Structured | | CoT | |
|---|---|---|---|---|---|---|
| | Standalone | w/ Reg. Filings (Ex. DeepSeek) | Standalone | w/ Reg. Filings (Ex. DeepSeek) | Standalone | w/ Reg. Filings (Ex. DeepSeek) |
| Excess Return | 0.004 | 0.010 (0.018) | 0.022 | 0.017 (0.027) | 0.030 | 0.020 (0.029) |
| Information Ratio | 0.06 | 0.20 (0.36) | 0.58 | 0.44 (0.69) | 0.68 | 0.37 (0.50) |
| Accuracy | 0.554 | 0.579 (0.590) | 0.579 | 0.589 (0.589) | 0.571 | 0.571 (0.575) |
| F1-Score | 0.525 | 0.553 (0.561) | 0.543 | 0.557 (0.556) | 0.552 | 0.555 (0.560) |

[13] Excluding DeepSeek's results, the inclusion of regulatory filings increases monthly excess returns from 1.7% to 2.5%, Information Ratios from 0.38 to 0.52, classification accuracy from 0.574 to 0.585, and F1-scores from 0.541 to 0.559.

#### 4.3.5 Performance Over Time

To further assess whether LLMs can extract coherent investment signals, we divide the live evaluation window into two subperiods. This partition enables us to examine the consistency of model performance over time, evaluate the impact of updated architectures, and determine whether anomalous results can be attributed to earlier iterations of LLM systems.

Figure 1 reports performance metrics by subperiod for both standalone prompts and configurations incorporating regulatory filings. Overall, we observe clear and consistent improvements over time across all prompting strategies, suggesting that successive model updates have enhanced the predictive quality of LLM-generated investment signals.

The improvement is particularly relevant for DeepSeek, which previously lagged under both standalone prompting and integration of regulatory disclosures. For standalone queries, DeepSeek's performance improves across all evaluation metrics, with the largest gains observed in excess returns (from 0.64% to 3.2% in the second subperiod) and classification accuracy (0.526 to 0.585). Moreover, limitations in processing high-density regulatory attachments appear partially mitigated in recent model versions. Although excess returns remain below those achieved by leading platforms, they become positive across all prompting configurations. Notably, the Information Ratio for DeepSeek queries incorporating regulatory disclosures increases from −0.52 to 0.18 in the second subperiod, indicating a substantial improvement.

However, model updates do not eliminate systemic reasoning failures. Even in early 2026, recurring analytical and computational errors persist across platforms and prompting strategies. Such failures can generate materially incorrect outputs that, in turn, lead to adverse financial decisions. Consequently, although performance gains are meaningful, our findings underscore the ongoing need for robust validation and governance mechanisms. Structured prompting, contextual integration, high-integrity inputs, and continuous expert supervision are all essential components of reliable deployment. Absent these safeguards, current LLMs remain vulnerable to systematic failures that compromise the quality and integrity of financial decision-making.

**Figure 1: Aggregate Performance by Subperiod**

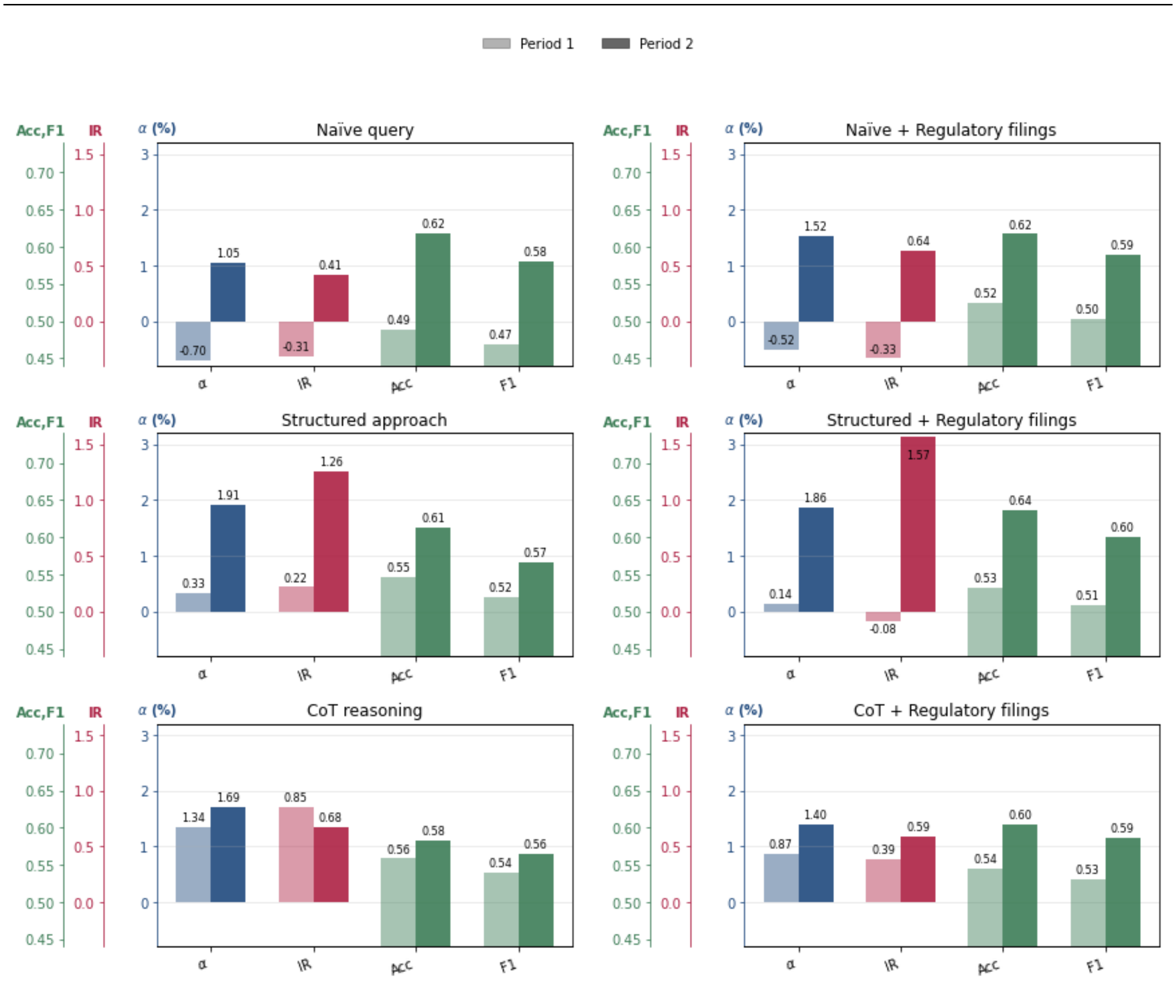

### 4.3.6 Sample Size and Software Configurations

The performance results presented in the preceding sections should be interpreted in light of two important constraints: the relatively short evaluation horizon—limited to ten monthly observations—and the rapidly evolving nature of large language models. Over our sample period (April 2025 to January 2026), model providers introduced frequent updates to underlying architectures, browsing capabilities, retrieval mechanisms, and safety constraints. Such modifications affect both model's reasoning behaviour and the information available at each evaluation date, complicating homogeneous comparisons.

Because short time horizons limit the statistical power of conventional inference, caution is warranted when interpreting point estimates. Accordingly, our evaluation adopts a holistic perspective that emphasizes robustness across prompting strategies, consistency across models and subperiods, and coherence between statistical and economic performance, rather than relying on individual figures.

These limitations, however, do not undermine the contribution of the study. To our knowledge, a ten-month live deployment in real-time financial markets constitutes the longest evaluation of LLM-based investment strategies to date. Moreover, our objective is not to establish definitive performance estimates or unconditional claims of model skill. Instead, we seek to evaluate whether—and under what conditions—LLM-generated signals can produce meaningful and reliable returns in a live-market setting. In this regard, the evidence presented in this study provides significant insights into the conditions needed to obtain coherent LLM outputs and the safeguards required for effective LLM deployments.

## 5. Conclusion

This paper evaluates whether large language models can generate economically reliable stock market predictions. By systematically comparing models, prompting strategies, and information sources, we examine whether—and under what conditions—LLM-generated signals can deliver risk-adjusted returns in excess of passive benchmarks.

Our findings demonstrate that LLMs can outperform the market under current technological capabilities. However, such outperformance is neither automatic nor robust in autonomous deployments. Recurring reasoning failures—spanning data retrieval, financial interpretation, computational processing, and meta-reasoning—persist throughout our live evaluation period. When left uncorrected, these errors propagate into investment decisions, generating unstable risk-adjusted performance and economically significant mistakes.

Live financial markets provide a particularly demanding test of LLM capabilities, requiring dynamic information retrieval, multi-step reasoning, numerical precision, and domain-specific judgment. Under simple conversational prompts, LLM-based recommendations are weakly informative: excess returns are inconsistent, Information Ratios are low and volatile, and classification accuracy only marginally exceeds random benchmarks. These limitations are especially relevant for retail or uninformed investors, who may lack the expertise required to detect subtle hallucinations, financial misconceptions, or numerically inconsistent outputs.

In contrast, structured prompting frameworks—and specifically supervised chain-of-thought protocols—systematically mitigate logical inconsistencies and reasoning noise. When LLMs are anchored in explicit analytical structures and subjected to iterative human validation, both economic performance and predictive accuracy materially improve. Hence, we find that converting raw generative capacity into reliable performance requires both structured prompting and human oversight.

We further show that incorporating regulatory filings enhances forecasting accuracy. Because supervisory disclosures contain standardized, high-integrity information, they serve as a grounding mechanism that facilitates cross-firm comparison and reduces informational noise. This finding is particularly relevant given the increasing role of AI in finance. In contrast to open-web content—which often contains contradictory narratives, biased interpretations, or factual inaccuracies—regulatory disclosures support more coherent reasoning and provides a reliable empirical anchor.

Looking ahead, our findings indicate that implementation challenges in AI-driven finance lie in integrating raw LLM capabilities into governance frameworks that combine analytical guidance, verified inputs, and human oversight. The challenge is not merely technical but organizational. Raw model capabilities do not seem enough to achieve reliable and robust results. Therefore, LLM architectures should be complemented by rigorous implementation safeguards, fostering a collaborative Human–AI framework that harnesses generative power while systematically controlling reasoning failures and model risk.

## Annex 1. LLM Versions over the Evaluation Period

This annex reports the LLM versions used throughout the evaluation period. Consistent with the real-time implementation of the study, model versions correspond to those available at each evaluation date. Figure 2 summarizes the evolution of model versions across platforms over time.

**Figure 2: Evolution of LLM Versions during the Evaluation Period**

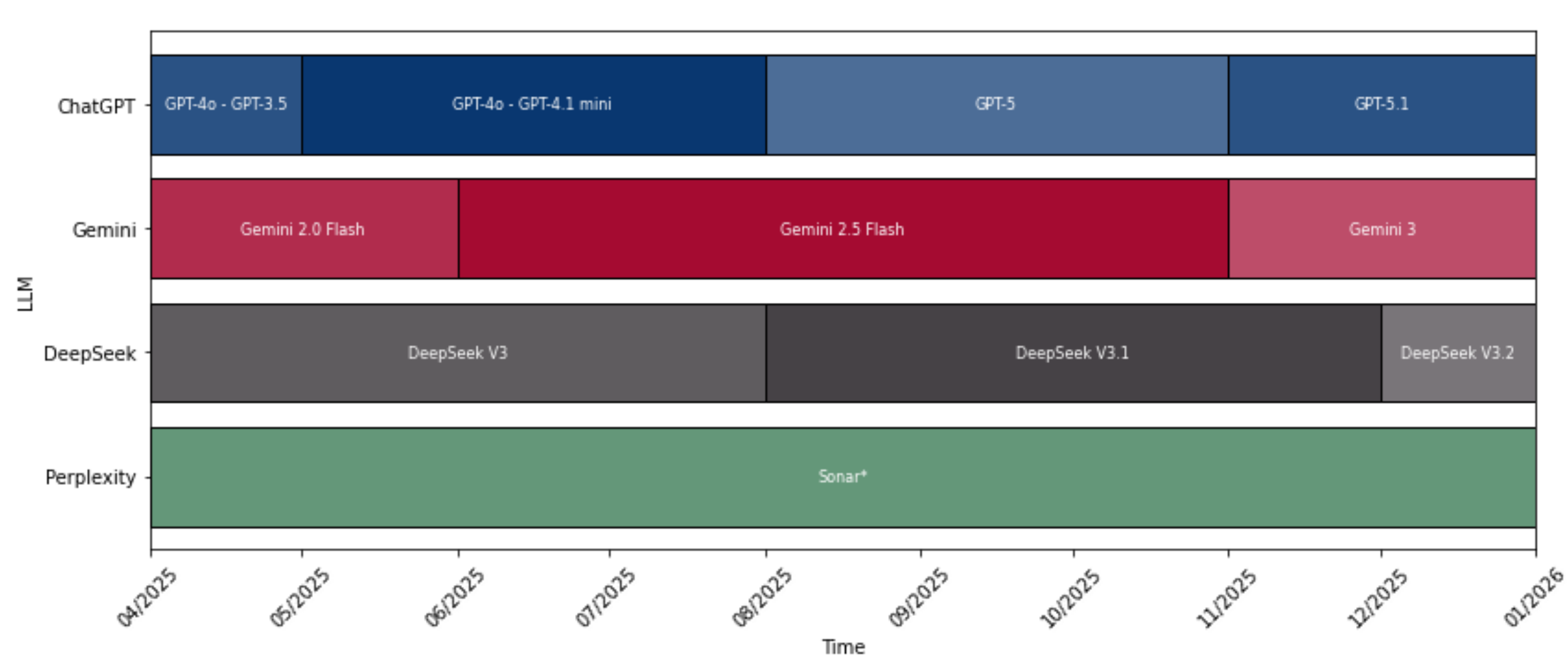

* Perplexity queries are conducted using the Sonar system. The platform does not disclose the specific underlying model or version associated with each response, as queries may be dynamically routed across different models.

# Annex 2. Prompting Templates

## 1. Naïve query

**Prompt design:** "Your role is a financial manager. I want to start investing in Spanish equities. Tell me which IBEX-35 assets you expect to outperform the stock market over the next month. You must calculate a score from 0 to 1, based on your stock prediction, for all IBEX-35 index components.

The score must depend on different categories and factors and reflect the expected return that you have calculated for each company using your model. You must include in your answer: 1. The overall score calculated for [firm X, firm Y and firm Z]. 2. A table for one of the assets showing: (i) the categories used in the calculation, (ii) the variables included in each category, (iii) the calculation formula for each variable, (iv) the raw value obtained for each variable and (v) the overall score calculated for the asset.

Check that in your response you have shown both the scores and the table."

## 2. Structured approach

**Prompt design:** "Today is [first trading day of the month] and the cutoff date is [last day of previous month]. Do not use information published after that day. Your role is a financial manager. I want to start investing in Spanish equities. Tell me which IBEX-35 assets you expect to outperform the stock market over the next month. You must calculate a score from 0 to 1, based on your stock prediction, for all IBEX-35 index components.

You must use the most recent data available to calculate each variable and consider forward-looking information and recent trends when available. The score must be calculated using a multi-factor model that includes 6 categories with the following weights and variables: 1. Valuation (20%): P/E ratio (60%) and P/B ratio (40%) 2. Growth potential (20%): EPS Growth (60%) and Revenue Growth (40%) 3. Financial health (15%): Debt/Equity Ratio (60%) and ROE (40%) 4. Technical (15%): Momentum (60%) and Relative Strength Index (RSI) (40%) 5. Macro & Sector risk (15%): Industry growth rate (60%) and Sector Outlook (40%) 6. Sentiment (15%): Recent news sentiment (1 month) (80%) and analyst's recommendations (20%). Use the general version of each formula and a [0,1] normalization range based on the typical ibex 35 values for each variable.

You must include in your answer: 1. The overall [0,1] score calculated for [Firm X, Firm Y and Firm Z]. 2. A table for one of the assets including: (i) the categories used in the calculation, (ii) the variables included in each category, (iii) the calculation formula for each variable, (iv) the raw values obtained for each variable, (v) the date to which each component in the formula refers to, (vi) the data source, (vii) the values used in the normalization range and (viii) the overall score calculated for the asset.

Check that in your response you have shown both the scores and the table. I'm going to tip $1000 for a better solution! You will be penalized if the model is not well built."

### 3. Chain-of-though reasoning

CoT prompts iteratively review the scores generated under the structured approach, identifying and correcting inconsistencies in model's reasoning, including arithmetic errors and reliance on stale or outdated information. Each score is evaluated individually to ensure logical consistency and numerical accuracy.

### 4. Analysis of regulatory filings

**Prompt design:** "Today is [first trading day of the month] and the cutoff date is [last day of previous month]. Do not use information published after that day. Your role is a financial manager. I want to start investing in Spanish equities. Tell me which IBEX-35 assets you expect to outperform the stock market over the next month. Analyze in detail the documents attached and based on your analysis of these regulatory filings, you must calculate a score from 0 to 1 for [Firm X] given your stock prediction."